\documentclass[aps,pra,twocolumn, showpacs]{revtex4}
\usepackage{graphicx}
\usepackage{dcolumn}
\usepackage{bm}

\begin{document}
\preprint{preprint}
\title{Collective Sideband Cooling in an Optical Ring Cavity}
\author{Th. Els\"{a}sser}
\author{B. Nagorny}
\author{A. Hemmerich}
\email{hemmerich@physnet.uni-hamburg.de}
\affiliation{Institut f\"{u}r Laser--Physik, Universit\"{a}t Hamburg, 
Jungiusstrasse 9, D--20355 Hamburg, Germany}

\date{\today}

\begin{abstract}

We propose a cavity based laser cooling and trapping scheme, providing tight 
confinement and cooling to very low temperatures, without degradation at high 
particle densities. A bidirectionally pumped ring cavity builds up a resonantly enhanced 
optical standing wave which acts to confine polarizable particles in deep 
potential wells. The particle localization yields a coupling of the degenerate 
travelling wave modes via coherent photon redistribution. This induces a 
splitting of the cavity resonances with a high frequency component, that is tuned to the 
anti-Stokes Raman sideband of the particles oscillating in the potential 
wells, yielding cooling due to excess anti-Stokes scattering.
Tight confinement in the optical lattice together with the prediction, 
that more than $50 \%$ of the trapped particles can be cooled into the motional ground state,
promise high phase space densities.

\end{abstract}


\maketitle

Laser cooling is a powerful technique to obtain the lowest temperatures 
accessible in laboratories to date \cite{reviewLC}, which has paved the pathway for the 
formation of Bose-Einstein condensates of atomic gases \cite{And:95}. Unfortunately, 
laser cooling has so far been constrained to a limited number of atomic species 
prepared at relatively low densities below $10^{12} cm^{-3}$. The reason 
is that all laser cooling schemes have used spontaneous emission in one or 
the other way to optically dissipate motional energy. To enable repeated excitation 
and spontaneous emission cycles, the particle to be laser cooled after 
spontaneous decay needs to return to a state that can be re--excited by 
laser radiation. This is only possible for a small class of atoms (e.g. 
alkali metals) with a sufficiently simple structure of the ground state. 
Molecules, for example, with their typically complex rotational and 
vibrational structure, would afford an unfeasibly large number of 
excitation frequencies. At high particle densities pontaneous emission poses 
the additional problem of uncontrolled re-absorption of fluorescence photons, which do not 
provide the highly ordered characteristics of the exciting laser radiation 
and thus yield undesired heating. These deficiencies of 
coventional laser cooling have inspired a quest for new optical cooling schemes 
which rely on coherent scattering, resonantly enhanced by optical 
resonators.

The interaction of atoms with a light mode of a high finesse cavity has
been a subject of extensive research in the past.  For long, most work has
focused on very small cavities with high values of the electric field per
photon and a few photons coupled to a few atoms.  Recent studies on such
cavities have also incorporated motional degrees of freedom
\cite{Pin:00, Hoo:00}.  In the past few years, high finesse cavities with larger
mode volumes have attracted interest as a possible means to cool
large atomic samples without the need for spontaneous photons. 
\cite{Doh:97, Hor:97, Hec:98, Vul:00}.  In a typical scenario referred to
as cavity Doppler cooling \cite{Vul:00, Cha:03}, polarizable particles
placed inside a linear resonator are irradiated from the side by an
off--resonant light pulse, yielding elastic Rayleigh scattering with a
scattering rate into resonant cavity modes, which is significantly enhanced
over the free space value.  By tuning the cavity resonance slightly above
the frequency of the incident light pulse, energy is extracted from the
motional degrees of freedom.

Unfortunately, such cavity based cooling techniques impose significant 
technical challenges. Because the temperature limit 
scales with the cavity linewidth, for accessing low temperatures 
extremely sharp resonance conditions have to be maintained, although the 
number of photons inside the resonator, that carry the information, required 
for frequency control techniques, is tiny. Particles to be cooled have to be well confined 
inside the mode volume, which for technical reasons is usually 
limited. As proposed in ref. \cite{Vul:01} an additional confinement 
potential may be applied, posing the problem of a good matching with the mode 
volume. 

In this article we propose a simple and robust cavity cooling scheme, which
may provide a solution to the aforementioned obstacles, combining 
cooling to very low temperatures with tight confinement in
a single light field. A bidirectionally pumped ring cavity forms a resonantly 
enhanced intense optical standing which acts to confine polarizable particles. 
The particle localization yields a coupling of the degenerate counter--propagating
travelling wave modes via coherent photon redistribution \cite{Hem:99}.  
This leads to a splitting of the cavity resonance with a high frequency component, that 
can be tuned to the anti-Stokes Raman sideband of the particles 
oscillating in the potential wells. As a consequence scattering on the 
anti-Stokes Raman sideband is resonantly enhanced, exceeding Stokes 
scattering into resonator or free space modes, and thus vibrational energy is 
dissipated into the light field. Tight trap potentials in the optical lattice
together with the prediction, that more than $50 \%$ of the trapped
particles can be cooled into the motional ground state, promise high phase
space densities.

\begin{figure}
\includegraphics[scale=0.5]{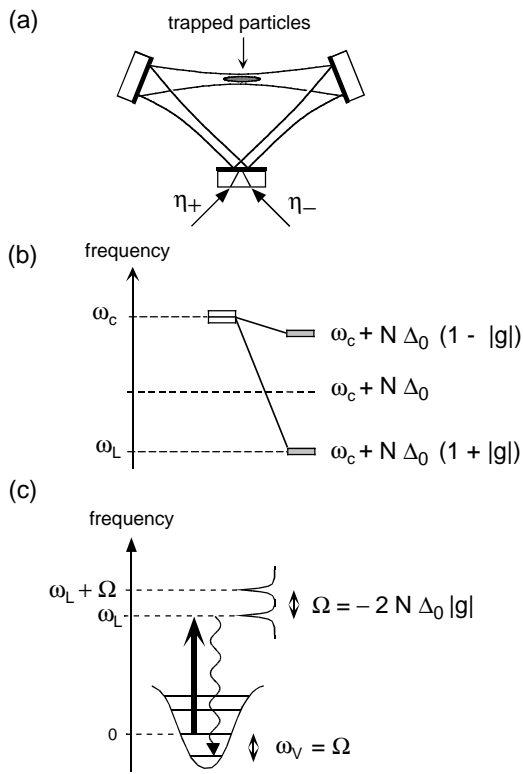}
\caption{ \label{Fig1} (a) Sketch of cooling scenario. 
(b) Sketch of mode splitting mechanism.
The travelling wave modes, indicated by the unfilled box in the center
are assumed to be resonant at $\omega_c$. When localized polarizable particles 
are present, the degeneracy is lifted and a resonance doublet arises (grey 
boxes on the right). 
(c) Energy budget of self-induced cavity sideband cooling. 
The lower frequency component of the resonance doublet in (b) provides
the optical lattice at frequency $\omega_L$, operating in
the Lamb-Dicke regime.  The higher frequency component $\omega_L + \Omega$
is tuned the anti-Stokes Raman sideband, i.e., $\Omega$ = $\omega_V$, where $\omega_V$ is the vibrational
frequency in the potential wells.}
\end{figure}

The cooling scenario is sketched in Fig.1(a). For simplicity, we refer to 
the particles to be cooled as atoms, bearing in mind, that nothing 
essential changes in the following discussion, if molecules were 
addressed. Laser light is resonantly coupled into both counter--propagating 
traveling wave modes of a ring cavity, which exhibits very high finesse 
(typically above $10^5$) and a large mode volume (typically a few $mm^3$). 
Inside the resonator an intense standing light wave is formed. Its 
frequency is tuned far away from resonant excitation frequencies of the 
atoms into a region of normal dispersion. The standing light field forms a 
one--dimensional optical lattice, i.e., steep light shift potentials can 
tightly trap the atoms well inside the Lamb-Dicke regime (given by 
$\omega_V \gg \omega_R$, where $\omega_V$ is the fundamental vibrational 
energy and $\omega_R$ is the single photon recoil energy) and the regime 
of resolved sidebands (given by $\omega_V \gg \gamma_c$ , where $\gamma_c$ 
is the intra--cavity field decay rate). If the sample of $N$ atoms is homogeneously
distributed, only forward scattering arises, i.e., the degenerate travelling 
wave modes are not coupled. Both modes remain eigenmodes, however shifted 
in frequency by an amount $N \Delta_0$, where $\Delta_0$ is
the lightshift per photon (see Fig.1(b)). This shift, which is 
negative for the relevant case of normal dispersion, is readily calculated by 
considering the refractive index of N atoms distributed over the mode volume.

The optical lattice acts to localize the 
atoms in the antinodes. The degree of localization may be described by
a parameter $g \equiv \frac{1}{N} \sum_{\nu = 1}^{N} e^{-i2kz_{\nu}}$ 
which takes values inside the complex unity sphere, where $z_{\nu}$ denote 
the deviations of the atomic positions from the adjacent potential 
minimum. For perfect localization, i.e., all atoms are positioned exactly in an 
anti--node ($z_{\nu}$ = 0), $g = 1$, while for a homogeneous atomic 
distribution $g = 0$. Generally, the complex phase of g scales with the 
center of mass and the modulus of g describes the spread of the atomic 
sample. In fact, for small deviations $kz_{\nu} \ll 1$, we may write  $g = 
1 - i 2k z_{cm}$, where $z_{cm}$ is the center of mass coordinate. If 
the phases $ kz_{\nu}$ are assumed to follow a Gaussian distribution, we 
may write $g = exp(-2k^2 \Delta z_{rms}^2) exp(-2ik z_{cm})$, where 
$\Delta z_{rms}$ denotes the root mean square spread of the atomic 
positions. 

For localized atoms additional back--scattering arises which couples the counter--propagating
travelling wave modes and lifts their degeneracy. In fact, two modified eigenmodes with 
orthogonal standing wave geometries arise. External pumping populates only one of the modes
which provides the lattice. This mode suffers an increased 
negative shift $N \Delta_0 \, \left(1 + |g| \right)$ 
because the atoms are localized in the antinodes, where the coupling 
to the light field is maximal. The other mode is not 
externally pumped and its nodes coincide with the center of mass positions of the trapped 
atoms. Thus, the coupling is minimized, yielding a decreased negative shift
$N \Delta_0 \, \left(1 - |g| \right)$. This less shifted eigenmode 
has a resonance frequency on the blue side of that of the 
optical lattice, i.e., emission into this mode dissipates energy into the 
light field. To get enhanced emission into this mode it may be tuned to 
coincide with the blue anti-Stokes sideband corresponding to the atoms trapped inside the 
optical potential wells, i.e., setting $\omega_V = 2 N 
|\Delta_0| \, |g|$ in Fig.1(c). This is readily achieved by adjusting $\omega_V$ 
via the intra-cavity light intensity.

The rate for resonant scattering into a resonator mode is 
$\eta_c \tilde \Gamma$, where  $\eta_c = 12F/\pi(k w_0)^2$  ($F$ = Finesse, 
$w_0$ = $e^{-2}$ radius of cavity mode) denotes the ratio of the scattering rate 
into a resonator mode to the free space scattering rate $\tilde \Gamma$.
For particles confined in the Lamb-Dicke regime scattering aquires a discrete 
frequency spectrum with an elastic component, which preserves the atomic 
motion, and modulation sidebands shifted by trap oscillation frequencies. The sideband 
intensities are suppressed by Lamb-Dicke factors which account for the 
spatial mismatch resulting from the change of the motional quantum state 
\cite{Win:79}. In the harmonic approximation, the suppression of scattering into the 
cavity modes is $\eta_{LD} = \omega_R/\omega_V$, while scattering into all other 
modes is suppressed by $\frac{2}{5} \, \eta_{LD}$. The extra geometry factor 2/5 
accounts for the angular distribution of the emitted k-vectors for 
scattering into free space modes. The rate for a resonant inelastic Raman 
scattering event  $|n \rangle \rightarrow |n-1 \rangle$, reducing the number of 
vibrational quanta by one, is thus given by  
$\Gamma_n = n \, \eta_{LD} \, \left( \eta_c + \frac{2}{5} \right) \, \tilde \Gamma$.
Raman scattering on the red Stokes sideband $|n-1 \rangle \rightarrow |n \rangle$ is not 
resonant with the cavity, i.e., for scattering into the cavity mode an 
extra suppression factor $\xi = (1+(2 \, \omega_V / \gamma_c)^2)^{-1}$ 
results from the cavity resonance profile. In the sideband regime
$\xi$ is much smaller than one. The corresponding rate for 
$|n-1 \rangle \rightarrow |n \rangle$ 
transitions is 
$\gamma_n = n \, \eta_{LD} \, \left( \xi \, \eta_c + \frac{2}{5} 
\right) \, \tilde \Gamma$. Cooling is expected, if $\gamma_n$ is significantly 
smaller than $\Gamma_n$, indicating that values of $\eta_c$ larger than the geometry 
factor 2/5 are required.

Solving the rate equations for the populations $\Pi_n$ of the 
vibrational levels $|n \rangle$, connected by the rates $\Gamma_n$ and 
$\gamma_n$, yields a steady state with 
$\Pi_n / \Pi_{n-1} = \left( \xi \, \eta_c + \frac{2}{5} \right) /
\left( \eta_c + \frac{2}{5} \right)$,
a corresponding population of the vibrational ground state of 
$\Pi_0 = \left( 1 - \xi \right) \eta_c / \left( \eta_c + \frac{2}{5} \right)$, 
and a mean vibrational quantum number 
$\langle n \rangle = \left( \eta_c \, \xi + \frac{2}{5} \right) / \left( 1 - \xi \right) \eta_c $.
For high finesse cavities we can realize values of 
$\eta_c$ on the order of one and $\xi$ on the order of $10^{-2}$, 
i.e., ground state populations above $50 \%$ and mean vibrational quantum 
numbers below 0.5 should be readily accessible. 
Calculating the total change of kinetic energy 

\begin{equation}
\frac{d}{dt} W = - \sum_{n = 0}^{\infty} \left(\Gamma_n - \gamma_{n+1} \right) \, \Pi_n
\end{equation}

in the classical limit ($\hbar \omega_V \ll k_B T$) yields exponential cooling according to

\begin{equation} 
dT/dt = - 2 \, \left( 1 - \xi \right) \, \eta_{LD} \, \eta_c \, \tilde \Gamma \, \, T \,
+ \left( \eta_c\, \xi + \frac{2}{5}\right) \tilde \Gamma \, T_R  ,
\end{equation}

with a cooling rate $2 \, \left( 1 - \xi \right)  \, \eta_{LD} \, \eta_c \, 
\tilde \Gamma$.
If $\xi \ll 1$, the main contribution to the temperature limiting second term on the right hand side
involving the recoil temperature $T_R$ results from scattering into free space.

To substantiate the results of the simple physical picture presented 
so far, a theoretical treatment starting from the dynamical equations of the 
system is required. In particular, a quantitative account of the mode coupling introduced by 
the atoms and the corresponding resonance condition should be produced.
Furthermore, the potential role of cavity mediated collective interactions 
needs to be clarified. In the following we take a semiclassical viewpoint, 
where the cavity field and the atomic motion is treated classically, while 
the atomic polarizability is taken from quantum mechanics. We start from 
the dynamic equations for the complex field amplitudes of 
the two degenerate modes of the empty cavity $\alpha_{\pm}(t)(z,t) = 
\alpha_{\pm}(t) exp(±ikz)$ ($k$ = wave number). According to ref. \cite{Gan:00} 
these equations write in the limit of low saturation and large detunings as

\begin{eqnarray}
\frac{d}{dt}
\left( \begin{array}{ccc} \alpha_+ \\ \alpha_- \\ \end{array} \right)
=  
\bf{M}
\left( \begin{array}{ccc} \alpha_+ \\ \alpha_- \\ \end{array} \right)
+
\gamma_0
\left( \begin{array}{ccc} \eta_+ \\ \eta_- \\ \end{array} \right)
\\
\bf{M} \equiv 
\left( \begin{array}{ccc} 
i \left( \delta_c - N \Delta_0 \right) - \gamma_c & - i N \Delta_0 \, g \\ 
 - i N \Delta_0 \, g^* &  i \left( \delta_c - N \Delta_0 \right) - \gamma_c \\ 
\end{array} \right)
\\
g \equiv \frac{1}{N} \sum_{\nu = 1}^{N} e^{-i2kz_{\nu}} \,\, ,
\end{eqnarray}

where $\delta_c$ is the detuning of the incoupled frequency from the 
resonance frequency of the empty cavity, 
$\gamma_0 = c/L$ ($c$ = speed of light, $L$ = cavity 
roundtrip length) is the free spectral range and $\eta_+$, $\eta_-$ are 
the complex field amplitudes of the incoupled light beams (all field 
amplitudes are scaled to the field per photon).

In order to describe cooling dynamics, we are looking for solutions of 
eq.3, which account for the presence of modulation sidebands. As already 
mentioned, such sidebands are expected to arise in the Lamb-Dicke regime, 
because the atoms oscillate in their potential wells with some frequency 
$\Omega$. We introduce the ansatz  
$\alpha_{\pm}(t)$ = $\alpha_{\pm} + \beta_{\Omega\pm} exp(-i\Omega t) + 
\gamma_{\Omega\pm} exp(i\Omega 
t)$ and $g = |g| (1 + i \epsilon \, cos(\Omega t))$ into eq.3, assuming 
symmetric pumping $\eta_+= \eta_- = \eta$ and find
							
\begin{eqnarray}
\alpha_{\pm} = \frac{\gamma_0 \, \eta}{\gamma_c - i\left( \delta_c - N \Delta_0 \, (1 + |g|) \right) }
\\
\beta_{\Omega\pm} = \gamma_{-\Omega\pm} =  
\frac{\alpha_{\pm} \, N \Delta_0 \, \epsilon}
{\gamma_c - i\left( \delta_c + \Omega - N \Delta_0 \, (1 - |g|) \right) }
\end{eqnarray}

The small phase parameter $\epsilon = 2k z_{cm,0}$ measures the maximum 
amplitude $z_{cm,0}$ of the center of mass oscillation of the sample.
For the carrier fields $\alpha_{\pm}$ forming the lattice, the resonance 
condition, obtained from eq.6 is $\delta_c = N \Delta_0 (1 + |g|)$, 
i.e., the detuning $\delta_c$ acquires a negative value ($\Delta_0 < 
0$, for normal dispersion). We set $\alpha_{\pm}$ to its resonant values 
$\gamma_0 \eta / \gamma_c$ in the following. 
The resonance condition for $\beta_{\Omega \pm}$ reads $\Omega = - 2 N \Delta_0 |g|$ (eq.7), 
and the corresponding frequency detuning of the empty cavity is $\delta_c + \Omega 
= N \Delta_0 (1 - |g|)$. Obviously, $\delta_c + \Omega > \delta_c$, 
showing, that at resonance the field 
amplitudes $\beta_{\Omega \pm}$ correspond to a blue detuned modulation sideband.
Similarly, the resonance condition for 
$\gamma_{\Omega \pm}$ is $\Omega = 2 N \Delta_0 |g|$ (eq.7), and the 
corresponding frequency detuning of the empty cavity is $\delta_c - \Omega 
= N \Delta_0 (1 - |g|) > \delta_c$. In this case the field 
amplitudes $\gamma_{\Omega \pm}$ correspond to a blue detuned resonant modulation sideband.
Obviously, either choice yields a resonant sideband at blue detuning, in 
accordance with our simple model (Fig.1(b)), 
i.e., emission on this sideband extracts kinetic energy from the system. 

We are now in the position to calculate the change of kinetic energy per 
particle $W$ by means of emission of photons on the high frequency 
sideband. This energy change is given by 

\begin{equation}
dW/dt = - \frac{2}{N} \, \frac{\Omega}{\omega_L} \, T_{loss} \, 
(P_{\beta}-P_{\gamma}) \, ,
\end{equation}

where $P_{\beta}$ and 
$P_{\gamma}$ are the powers in the traveling waves $\beta_{\Omega \pm}$ and $\gamma_{\Omega \pm}$ 
connected with the blue detuned resonant modulation sideband and the red detuned 
off-resonant counterpart respectively, and 
$T_{loss}$ is the total loss from the resonator given by $2 \gamma_c/ 
\gamma_0$. Eq.8 is evaluated as follows.
We account for the fact that $\Omega$ is determined by the 
lattice potential wells by setting $\Omega = \omega_V$ in the following.
We use eq.7 for the case of resonance for $\alpha_{\pm}$ to express
$P_{\beta}$ and $P_{\gamma}$ by the power $P_{\alpha}$ in each of the travelling 
wave modes $\alpha_{\pm}$, which can be further expressed by the 
free space scattering rate $\tilde \Gamma$.
Upon the assumption of a Gaussian ensemble, the square of $\epsilon$
is expressed as $\epsilon^2 = 4k^2 \Delta z_{rms}^2/N$. The 
relations $2W = k_B T = m \Delta z_{rms}^2 \omega_V^2$ (m = particle mass)
are used to finally obtain

\begin{eqnarray}
\frac{d}{dt} T =  - 2 \, \eta_c \, \eta_{LD} \, \tilde \Gamma \, 
\left(\pounds(\omega_V) - \pounds(-\omega_V) \right)
\\
\pounds(\omega_V) \equiv
\frac{\gamma_c^2}{\gamma_c^2 + \left( 2 N \Delta_0 \,|g| + \omega_V \right)^2} \, ,
\end{eqnarray}		
														
which reproduces the exponential cooling dynamics with
the same cooling rate as found in eq.2. Note that at resonance 
$\omega_V = -2 N \Delta_0 |g|$ the term in the brackets on the right hand side of eq.9 
becomes $1 - \xi$. The second (temperature limiting) term of eq.2 is naturally not reproduced, 
since free space scattering is not accounted for in eq.9. Our 
dynamical derivation of the cooling rate reveals the important role of the 
center of mass motion which is damped with a rate proportional to the number 
of particles $N$. The relative motional degrees of freedom are only damped 
via their small statistical coupling to the center of mass motion scaling 
with the inverse of $N$. Our calculation confirms, that the resonance splitting, which is 
essential for the cooling mechanism, results from the common action of all 
atoms.

Finally, we briefly sketch a realistic scenario, to experimentally test 
the new cooling scheme with rubidium atoms. We assume a triangular ring 
cavity similar to that reported in ref.\cite{Nag:03} with a finesse of 
$1.8 \times 10^5$, a bandwidth  $\gamma_c = \pi \, 17$~kHz and a mode radius $w_0 = 
130 \, \mu m$, i.e. $\eta_c \approx 0.6$. At 0.1~nm red detuning of the lattice 
frequency with respect to the D2 line at 780.24~nm and with 50~mW power 
circulating in each propagation direction the free space scattering rate 
becomes $\tilde \Gamma \approx 8 \times 10^3 s^{-1}$ and the Lamb-Dicke parameter is 
$\eta_{LD} \approx 10^{-2}$. The corresponding cooling rate is $2 
\eta_{LD} \, \eta_c \, \tilde \Gamma \approx 100 s^{-1}$, i.e., if the resonance
condition $\omega_V = - 2 N \, \Delta_0 |g|$ is maintained, the temperature decreases 
by 1/e in 10~ms. The sample is cooled to a mean vibrational quantum number of
$\langle n \rangle \approx 0.66$ which corresponds to a kinetic 
temperature of approximately 12~$\mu K$.  
Since the light shift per photon is $\Delta_0 \approx 1 \, s^{-1}$ and the 
vibrational frequency is $\omega_V \approx 2 \pi \, 380$~kHz (corresponding 
to a trap depth of $460 \, \mu K$), the resonance 
condition $\omega_V = - 2 N \, \Delta_0 |g|$ is satisfied with $N \approx 1.3 
\times 10^6$ atoms and a value for the localization of $|g|=0.9$. 
According to ref. \cite{Nag:03} this can be realized with no difficulties by 
loading the lattice with a standard magneto-optic trap yielding initial 
phase space densities of several $10^{-6}$ at a temperature of about $120 \, \mu K$. 

In summary, we have proposed a novel laser cooling scheme, which can provide 
tight confinement and nearly zero vibrational temperature in an optical 
lattice operating in the Lamb-Dicke regime. Because this scheme relies on coherent 
Raman scattering, it is applicable to any polarizable particle and does 
not suffer degradation at high particle densities. In this 
article we have discussed a one-dimensional scenario, however, an 
extension to two and three dimensions should not present any difficulties. 
A particularly simple way to add cooling in the transverse lattice planes is to 
combine our scheme with cavity Doppler cooling.

\begin{acknowledgments}
This work has been supported by Deutsche For\-schungs\-gemeinschaft (DFG) under 
contract number $He2334/3-2$. A.H. thanks Claus Zimmermann and Helmut 
Ritsch for many vivid discussions.
\end{acknowledgments}

\end{document}